# In-gap states and strain-tuned band convergence in layered structure trivalent iridate $K_{0.75}Na_{0.25}IrO_2$

Xujia Gong,[1] Carmine Autieri,[2] Huanfu Zhou,[3] Jiafeng Ma,[1] Xin Tang,[3] Xiaojun Zheng,[1] Xing Ming,[1*]

*1. College of Science, Guilin University of Technology, Guilin 541004, People's Republic of China*

*2. International Research Centre Magtop, Polish Academy of Sciences, Aleja Lotników 32/46, PL-02668 Warsaw*

*3. Key Lab of New Processing Technology for Nonferrous Metal & Materials, Ministry of Education, School of Materials Science and Engineering, Guilin University of Technology, Guilin 541004, China.*

## ABSTRACT

Iridium oxides (iridates) provide a good platform to study the delicate interplay between spin-orbit coupling (SOC) interactions, electron correlation effects, Hund's coupling and lattice degree of freedom. However, overwhelming investigations primarily focus on tetravalent ($Ir^{4+}$, $5d^5$) and pentavalent ($Ir^{5+}$, $5d^4$) iridates, far less attention has been paid to iridates with other valence states. Here, we pay our attention to a less-explored trivalent ($Ir^{3+}$, $5d^6$) iridates, $K_{0.75}Na_{0.25}IrO_2$, crystalizing in a triangular lattice with edge-sharing $IrO_6$ octahedra and alkali metal ions intercalated $[IrO_2]^-$ layers. We theoretically determine the preferred occupied positions of the alkali metal ions from energetic viewpoints and reproduce the experimentally observed semiconducting behavior and nonmagnetic (NM) properties. The SOC interactions play a critical role in the band dispersion, resulting in NM $J_{eff} = 0$ states. More intriguingly, our electronic structure not only uncovers the presence of in-gap states and explains the abnormal low activation energy in $K_{0.75}Na_{0.25}IrO_2$, but also predicts the band edge can be effectively modulated by mechanical strain. Especially, the in-gap states feature with enhanced band-convergence characteristics by 6% compressive strain, which will greatly enhance the electrical conductivity of $K_{0.75}Na_{0.25}IrO_2$. Present work sheds new lights on the unconventional electronic structures of the trivalent iridates, indicating its promising application as nanoelectronic and thermoelectric material.

---

[*]**Email:** mingxing@glut.edu.cn

## I. INTRODUCTION

Due to the intricate interplay between electron correlation, Hund exchange coupling, spin-orbit coupling (SOC) and crystal field splitting, iridium oxides (iridates) exhibit novel phenomena and intriguing properties [1-8]. Normally, the *d* orbitals are split into three-fold degenerate $t_{2g}$ and two-fold degenerate $e_g$ states by the octahedral crystal field. Interestingly, attributed to large octahedral crystal-field splitting and strong SOC interactions in iridates, the triply degenerate $t_{2g}$ states will further evolve into $j_{eff}$ = 3/2 quartet and $j_{eff}$ = 1/2 doublet states [9]. The $j_{eff}$ state has been proposed as a common ingredient in iridates, arousing researchers' passion to explore the interplay between electron correlations and SOC interactions. Depending on the relative strength of Coulomb repulsion and SOC interactions, exotic phases have been revealed in iridates [1,4], such as spin-orbit coupled Mott insulators [9,10], Weyl semimetals [11,12], topological insulators [4,13], giant magnetic anisotropy [14,15], superconductors [16-19], spin liquids and spin ices [20,21]

Furthermore, the effective electron correlations and SOC interactions increase with the decreasing geometry connectivity (the connection manner between the polyhedral units) in iridates [5-7]. The crucial roles played by the lattice degree of freedom and geometric connectivity in iridates have renewed growing interests in iridates. The $IrO_6$ octahedra often arrange in a mixed corner- and edge-sharing network. The structural dimensionality and the stacking manner of the $IrO_6$ octahedra can be used to manipulate the electronic structure of iridates [6,22], such as dimensionality-driven metal-insulator transition (MIT) in the Ruddlesden-Popper series $Sr_{n+1}Ir_nO_{3n+1}$ compound [5,23], novel $j_{eff}$ = 1/2 Mott insulating state in quasi-two-dimensional $Sr_2IrO_4$ [9,10], and semimetallic $SrIrO_3$ with unusually narrow bandwidths [24]. Overwhelming investigations have focused on the tetravalent and pentavalent iridates with $5d^5$ ($Ir^{4+}$) and $5d^4$ ($Ir^{5+}$) electronic configurations, far less attention have been paid to iridates with other valence states and electronic configurations.

Recently, an unprecedented trivalent iridate $K_{0.75}Na_{0.25}IrO_2$ was synthesized by Weber *et al.*, where the $Ir^{3+}$ ions adopting a $5d^6$ electron configuration [25]. As shown in **Figure 1**, $K_{0.75}Na_{0.25}IrO_2$ crystallizes in a triangular lattice with *P*6$_3$/mmc space group. The edge-sharing $IrO_6$ octahedra form two-dimensional $[IrO_2]^-$ layers. The alkali metal ions act as intercalated ions, occupying two types of interlayer sites: A (1/3, 2/3, 1/4) and B (0, 0, 1/4), respectively. Experiments indicate that the Na ions almost only occupy the interlayer positions A, whereas two third (one third) of the K ions occupy the interlayer positions A (B), concomitant with 25% (75%) vacancies in the positions A (B).

The uncommon trivalent iridium ions and the triangular layered structure of $K_{0.75}Na_{0.25}IrO_2$ offer a promising platform to explore the interplay between electron correlations, SOC interactions, and lattice degree of freedom. $K_{0.75}Na_{0.25}IrO_2$ shows nonmagnetic (NM) characteristics and semiconducting behavior by magnetic property and electrical resistivity measurements. However, experimentally determined activation energy (85.7 meV) is very low for this semiconductor. Although elementary electronic structure has been attached in the originally experimental work, the authors only considered one composition of the alkali metal ions by completely neglecting the Na ions, where the interlayer positions only occupied by K ions. Therefore, the electronic structure corresponds to $KIrO_2$, showing a big band gap of 0.973 eV, and the abnormal activation energy remains to be clarified [25].

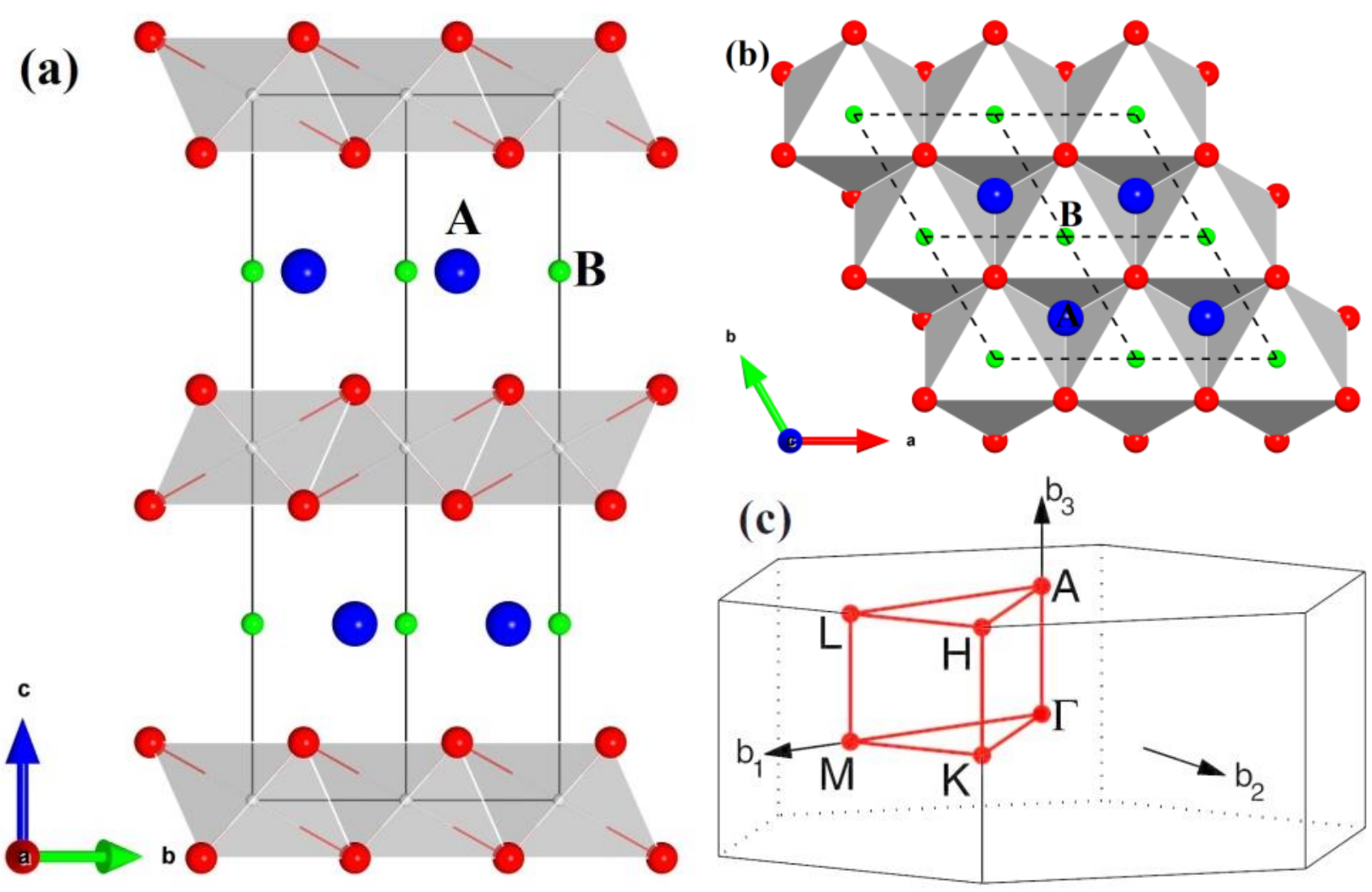

**Figure 1** (a) Side view and (b) top view of the crystal structure of $K_{0.75}Na_{0.25}IrO_2$ drawn with VESTA [26], and (c) corresponding *k*-path of the Brillouin zone generated with AFLOW [27,28]. The $[IrO_2]^-$ layers consist of edge-sharing $IrO_6$ octahedra on a triangular lattice (the dashed line denotes a unit cell). Two types of interlayer sites are occupied by the alkali metal ions, where the big/small balls denote the positions A/B facing hallow/face of the octahedra.

In the present work, we not only consider the interlayer positions fully occupied by one type of alkali metal ions, but also consider more complexed occupancies of the mixed alkali metal ions. Especially, in order to capture the real occupancy of the alkali metal ions and inspect the intrinsic characteristics of the title compound of $K_{0.75}Na_{0.25}IrO_2$, we have built supercell to consider the distribution of alkali metal ions and vacancies in the interlayer space. These settings of the structure model are much more consistent with the real composition of $K_{0.75}Na_{0.25}IrO_2$. Based on density functional theory (DFT) first-principles electronic structure calculations, we try to provide new insights into the unique electronic structures of this uncommon trivalent iridate. We not only successfully reproduce the experimentally observed semiconducting behavior and NM properties, but also reveal the evolutions of the band structure along with the SOC interactions. More interestingly, the in-gap states featured with nearly free electron characteristics are discovered below the conduction bands. Intriguingly, the electronic structure exhibit excellent tunability under mechanical strains, which pave the way for realization band convergence and multiple valleys in $K_{0.75}Na_{0.25}IrO_2$. The present work provides clues to explain the experimentally observed abnormally low activation energy and indicates its potential applications in nanoelectronic and thermoelectric fields.

## II. COMPUTATIONAL DETAILS

Employed the projector augmented wave (PAW) method [29,30], first-principles calculations were performed in the framework of DFT with VASP code [31], together with the Perdew-Becke-Ernzerhof (PBE) parameterization of the generalized gradient approximation (GGA) as the exchange-correlation functional [32]. We adopted the rotationally invariant DFT + $U$ method introduced by Liechtenstein *et al*. to consider the correlations effects [33]. The onsite Coulomb interactions $U$ and Hund coupling parameter $J_H$ were set to be 2 and 0.2 eV for the Ir atoms, respectively [5,14,15]. Calculations have also taken into account the fully relativistic SOC interactions. The total energy was converged within $10^{-6}$ eV. The Brillouin zone was sampled using a $k$-point mesh of 14 × 14 × 3 for the crystallographic unit cell, and the energy cutoff was set to 520 eV.

In order to reveal the preferred occupation positions of the alkali metal ions, first we considered ten configurations in the crystallographic unit cell (schematic models and detailed descriptions were presented in **Figure S1** in SM [34]), which are denoted as NaA, NaB, NaANaB, KA, KB, KAKB, KANaB, KBNaA, KANaA and KBNaB, respectively. Recognized the structural stability and experimental measured positions of the alkali metal ions, we did not consider the configuration that the alkali metal ions only occupy one interlayer position and left another interlayer position empty. Due to the limitation of the unit cell, real compositions of the model are $NaIrO_2$, $KIrO_2$ or $K_{0.5}Na_{0.5}IrO_2$ in these ten configurations. In order to maximize reflect the occupancy of alkali metal ions in real material of $K_{0.75}Na_{0.25}IrO_2$, we build 2×2×1 supercells and consider three different occupancy configurations of the alkali metal ions according to the experimentally determined crystallographic structure (as shown in **Figure 2**).

Regarding the layered structure and large interlayer distances along the *c*-axis, beyond traditional GGA, we took into account the van der Waals (vdW) interactions using different semiempirical dispersion interaction correction methods to relax the crystal structure. However, the vdW corrections play minor role in the lattice constants [see **Figure 3(a)** and **Table S1** in Supplemental Material (SM) [34] for a detailed comparison]. Therefore, we employed the semiempirical dispersion interaction correction methods (DFT-D) with optPBE-vdW functional [35-37] because they nearly predicted identical results.

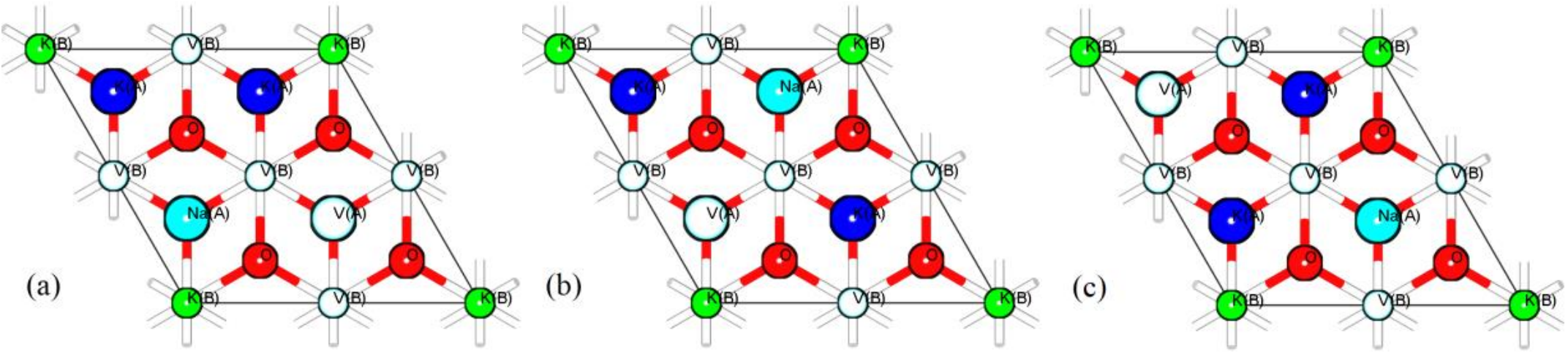

**Figure 2** Hypothetical three mixed occupancy configurations of alkali metal ions according to statistical distribution in the interlayer space of $K_{0.75}Na_{0.25}IrO_2$ iridate. The solid line denotes one 2×2×1 supercell, and there are three K ions and one Na ion in each supercell, corresponding to nominal composition of $K_{0.75}Na_{0.25}IrO_2$. 1/4 (2/4) A positions are occupied by Na (K) ions, whereas only 1/4 B positions have been occupied by K ions, resulting in 25%/75% vacancies in positions A/B marked as V(A) and V(B).

## III. RESULTS AND DISCUSSIONS

Firstly, we pay our attention to the preferential occupation of the alkali metal ions. As shown in **Figure 3(a)**, the in-plane lattice constant *a* is almost independent on the occupation positions of the alkali metal ions. In contrast, the out-of-plane lattice constant *c* obviously depends on the occupied positions of the alkali metal ions, which is longer when the alkali metal ions occupy the interlayer position B. As shown in **Figure 1(b)**, these trends can be attributed to the different local geometry of the two interlayer positions, where positions A face to two hollows of the $[IrO_2]^-$ layers, while positions B face to two surfaces of the $IrO_6$ octahedra. In addition, owing to the bigger ionic radius of K ions, the lattice constant *c* is obvious longer when the interlayer positions are only occupied by K ions. The experimentally observed preferred-occupancy of the alkali metal ions can be further inspected from the energetic point of view. As shown in **Figure 3(b)**, no matter there is one or two types of alkali metal ions, once the alkali metal ions occupy the interlayer positions A, the energies are obvious lower than they occupy the positions B, which are closely related to the hollows of the $[IrO_2]^-$ layers. Consequently, these two types of alkali metal ions prefer to occupy the interlayer positions A with much lower energy, which are in good agreement with the experimental results, where Na ions almost only occupy the positions A, whereas two third (one third) of K ions occupy positions A (B) [25].

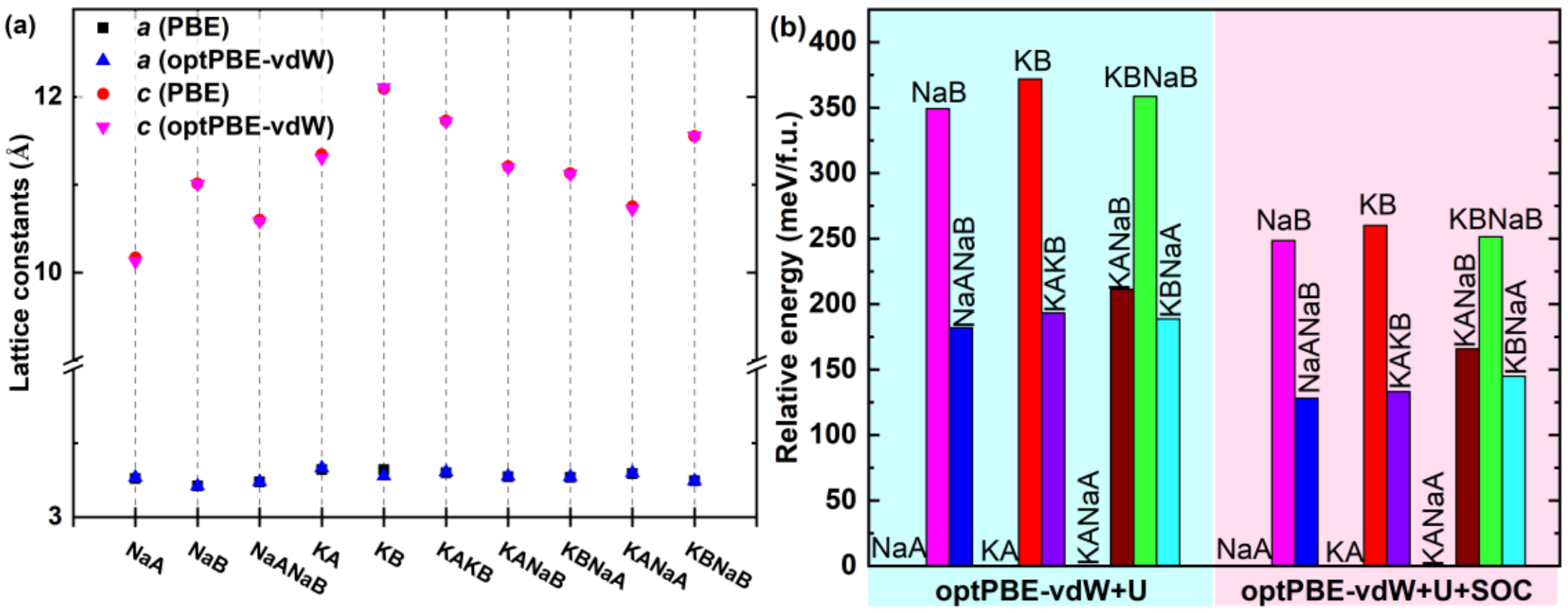

**Figure 3** (a) Theoretically calculated lattice constants with PBE or optPBE-vdW functionals and (b) relative energies calculated without or with SOC by optPBE-vdW + *U* calculations. The energies for the configurations of NaA, KA and KANaA are set as references (zero point).

The original DFT calculations solely consider the interlayer positions are only occupied with K ions by completely neglecting the Na ions, which is inconsistent with the occupancy and composition of the alkali metal ions in real material of $K_{0.75}Na_{0.25}IrO_2$ [25]. Obviously, the resultant electronic structure cannot express the inherent characteristics of $K_{0.75}Na_{0.25}IrO_2$. In this context, we resort to the 2×2×1 supercells and consider three different occupancy configurations of the alkali metal ions as shown in **Figure 2**. Despite the energy of the second case (**Figure 2(b)**) is significantly lower than the other two cases, the projected band structures show similar characteristics for these three cases (**Figure 4** and **Figures S2-S3** of SM [34], hereafter we just discuss and present the results of $K_{0.75}Na_{0.25}IrO_2$ corresponding to the occupancy configuration of **Figure 2(b)**. The spin polarized calculations are performed to check the basic electronic structure, and the electron correlation interactions together with vdW corrections are considered by optPBE-vdW + *U* calculations.

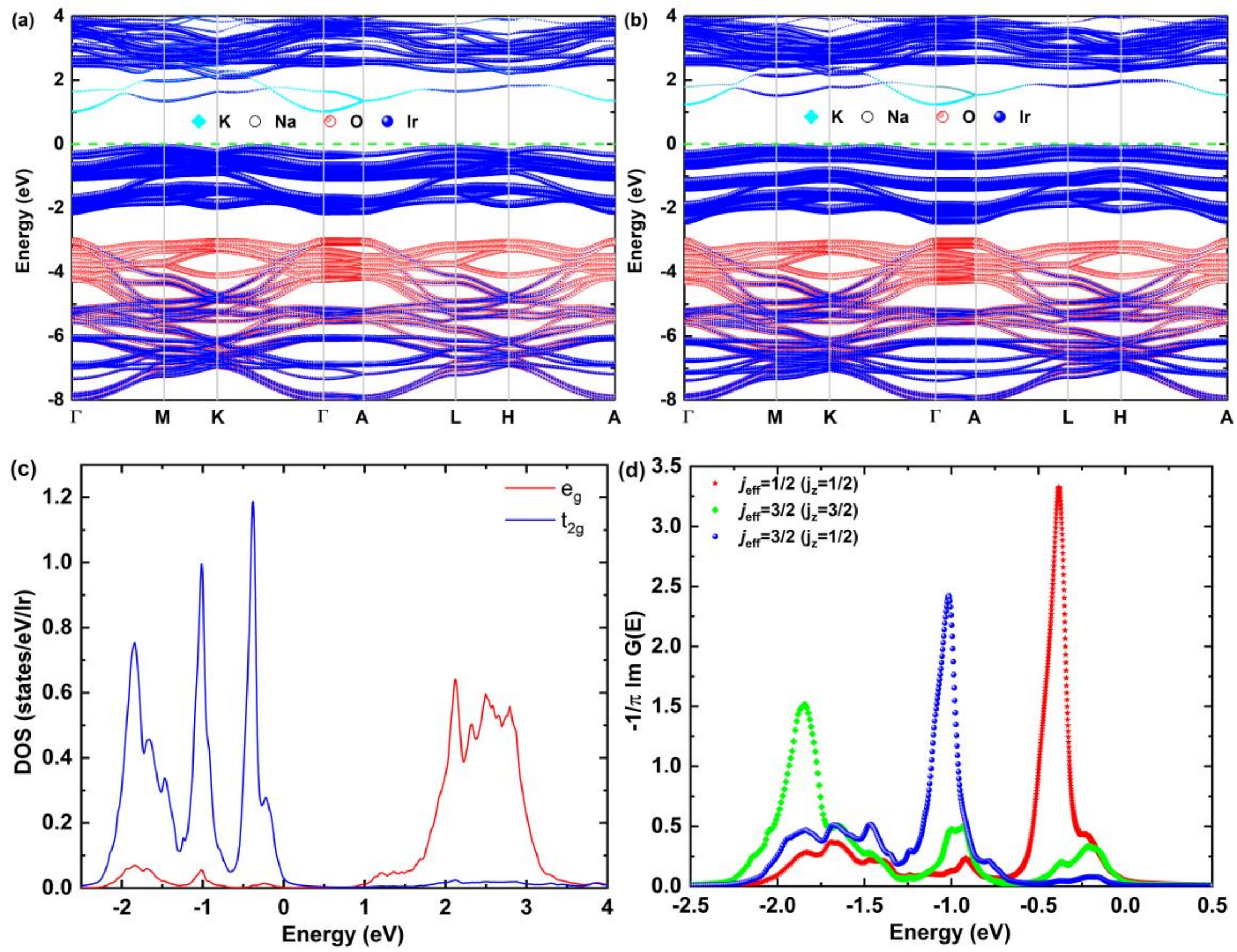

**Figure 4** Electronic structure of $K_{0.75}Na_{0.25}IrO_2$ with the mixed occupancy configurations of alkali metal ions as shown in **Figure 2**(b). (a) and (b) are projected band structures calculated within optPBE-vdW + *U* without SOC and including SOC, respectively. The bands are projected onto K 4*s*, Na 3*s*, O 2*p* and Ir 5*d* states, denoted by cyan diamond, black circle, red hollow balls and blue solid balls. The size of the symbols is proportional to the contribution from the corresponding elements. The projection of Ir 5*d* states onto the $t_{2g}$, $e_g$ states is presented in (c), and the corresponding projection of the $t_{2g}$ states onto the $j_{eff}$ states is plotted in (d).

As shown in **Figure 4**, the strong hybridizations between Ir 5*d* and O 2*p* states give rise to the Ir 5*d* bonding bands ranging from -8 to -5 eV [14,15,38-40]. The O 2*p* states mainly locate above these bonding states, whereas the Ir 5*d* antibonding states dominantly contribute to the states near $E_F$. The Ir 5*d* valence bands are almost flat without dispersion along the Γ-A direction (along the stacking direction of the $[IrO_2]^-$ layers in real space), reflecting weak interlayer interactions in these layered structure iridates [41]. One noteworthy feature is the Ir 5*d* orbitals have been split into the $t_{2g}$ and $e_g$ states by the octahedral crystal field. Even without onsite Coulomb corrections, the crystal-field splitting is sufficient to account for the insulating nature, the experimentally observed semiconducting behavior has been successfully reproduced. As shown in **Figures 4(a)** and **(c)**, the isolated manifold of $t_{2g}$ bands below Fermi level ($E_F$) are fully occupied, and are distinctly separated away from the fully empty $e_g$ states above $E_F$ by a large crystal-field splitting, resulting in insulating band structures. Due to the two-dimensional connectivity of the $IrO_6$ octahedra, the bandwidth of the $t_{2g}$ states (about 2 eV) are close to another quasi two-dimensional iridate $Sr_2IrO_4$ [5,9]. The band structure displays $5d^6$ ($t_{2g}^6$, $e_g^0$) electronic configuration with typical NM characteristics, coinciding well with the magnetic measurement results [25]. According to traditional band theory scenario, the layered iridate $K_{0.75}Na_{0.25}IrO_2$ can be classified as band insulator because of even-number of electrons [42,43].

To shed more lights on the underlying role of SOC in the electronic structure, the SOC interactions are included by optPBE-vdW + $U$ + SOC calculations. As shown in **Figure 4(b)**, SOC interactions significantly influence the dispersion of the Ir-$t_{2g}$ states below $E_F$, but almost have no impact on the dispersion of the Ir-$e_g$ states, the triply degenerate $t_{2g}$ states evolve into $j_{eff}$ = 3/2 quartet and $j_{eff}$ = 1/2 doublet states, which are all fully occupied by the six electrons of the trivalent $Ir^{3+}$ ($5d^6$) ions [1,9]. Because the energy of the $j_{eff}$ = 3/2 states is somewhat lower than that of the $j_{eff}$ = 1/2 states, there are obvious gaps between them [1,7,9]. To get further insight into the nature of the $j_{eff}$ states, we perform non-spin polarized calculations for the NM state of $K_{0.75}Na_{0.25}IrO_2$ using the projection-embedding implementation [44] on top of the WIEN2K package [45]. The projected density of states corresponds to the imaginary part of the Green's function ($-1/\pi \ Im \ G(E)$) [7]. As shown in **Figure 4(d)** and **Figure S4** of SM [34], the projection of the $t_{2g}$ states onto the $j_{eff}$ states is very convenient to identify the $j_{eff}$ = 1/2 and $j_{eff}$ = 3/2 components. Particularly, the $j_{eff}$ = 1/2 states show very high purity, which are distinctly separated from the quartet $j_{eff}$ = 3/2 states. In the strong SOC scenario, the fully occupied $j_{eff}$ states give rise to a NM $J_{eff}$ = 0 characteristics.

Another remarkable feature of the electronic structure of $K_{0.75}Na_{0.25}IrO_2$ is the in-gap states locating at the bottom of the conduction bands. The conduction band minimum (CBM) always located at the high symmetry point Γ, and the states around the high symmetry points Γ and A primarily come from the interlayer alkali metal ions, rather than the Ir 5$d$ states (detailed projected band structures of the CBM are displayed in **Figures 5-6**). Although previous experiment has proposed the existence of in-gap states caused by the extrinsic impurities or lattice defects [25], our band structures demonstrate the in-gap states originate from the intrinsic electronic states of the alkali metal ions. We propose that the insulating $[IrO_2]^-$ layers act as the blocking layer, while the alkali metal ions intercalate the interlayer space of the building block of the $[IrO_2]^-$ layers, and generate the in-gap state. Especially, these in-gap states near Γ point have nearly free electron (NFE) characteristics with parabolic energy dispersions [46]. By fitting the energy band in the vicinity of the Γ point, the effective masses m* along the Γ-A, Γ-K, and Γ-M directions are estimated to be 0.586, 0.493 and 0.557 $m_0$ ($m_0$ is the free-electron mass) for $K_{0.75}Na_{0.25}IrO_2$ calculated by optPBE-vdW + *U* without SOC. Including SOC interactions, the corresponding effective masses m* are 0.639, 0.513 and 0.603 $m_0$, respectively. The deviations of effective masses m* from free-electron mass $m_0$ widely exist in other NFE systems. For example, the m* of two NFE states in the bulk gap of silicon are 0.404 and 0.314 $m_0$ [47], and the m* of NFE states in carbon and boron nitride nanotubes are 1.2 $m_0$ [48]. The in-gap states and NFE characteristics provide clues to the experimentally observed abnormal low activation energy of $K_{0.75}Na_{0.25}IrO_2$.

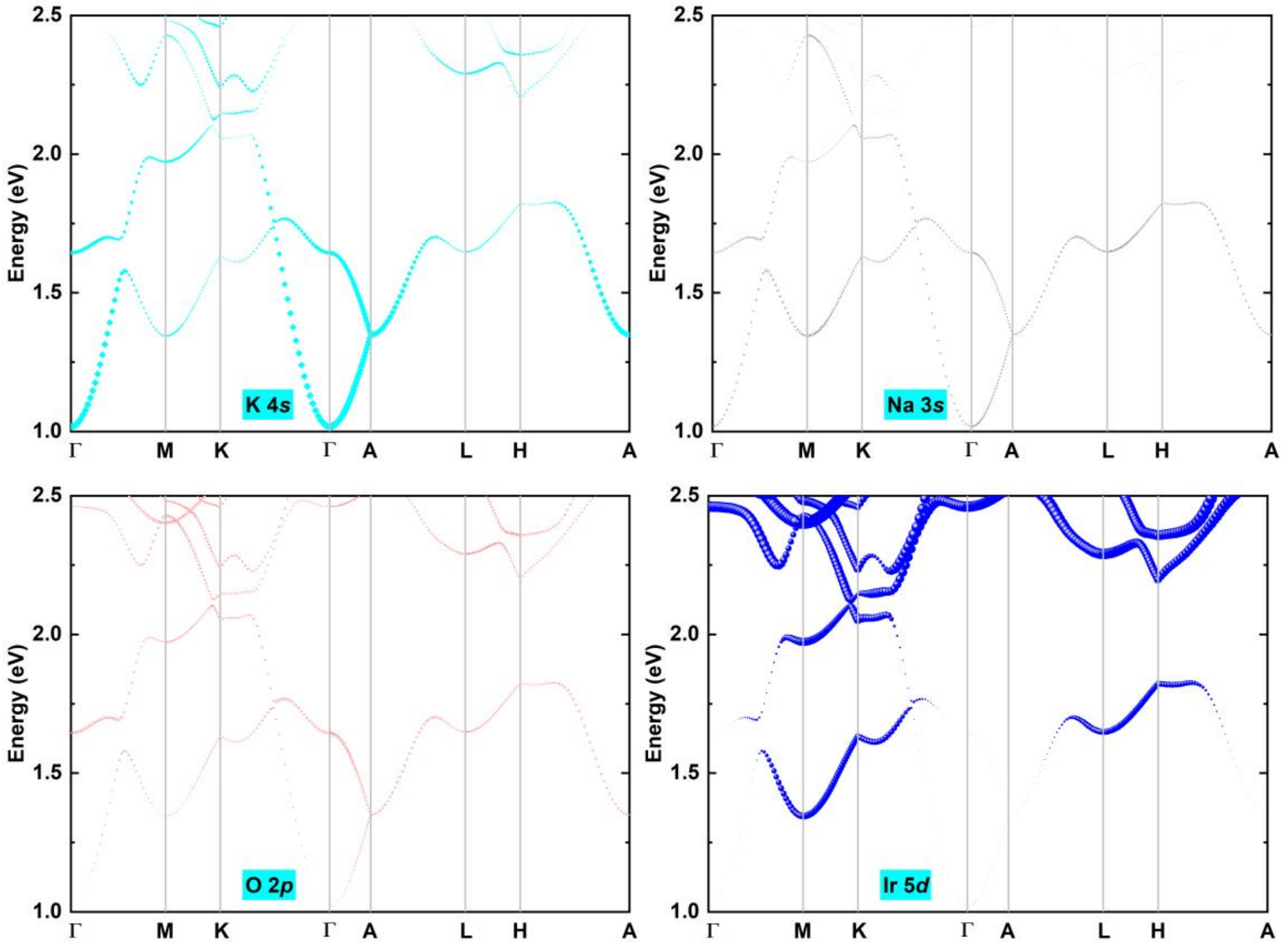

**Figure 5** Projected CBM of $K_{0.75}Na_{0.25}IrO_2$ calculated within optPBE-vdW + *U* without SOC, which are projected onto K 4*s*, Na 3*s*, O 2*p* and Ir 5*d* states, denoted by cyan diamond, black circle, red hollow balls and blue solid balls. The size of the symbols is proportional to the contribution from the corresponding elements.

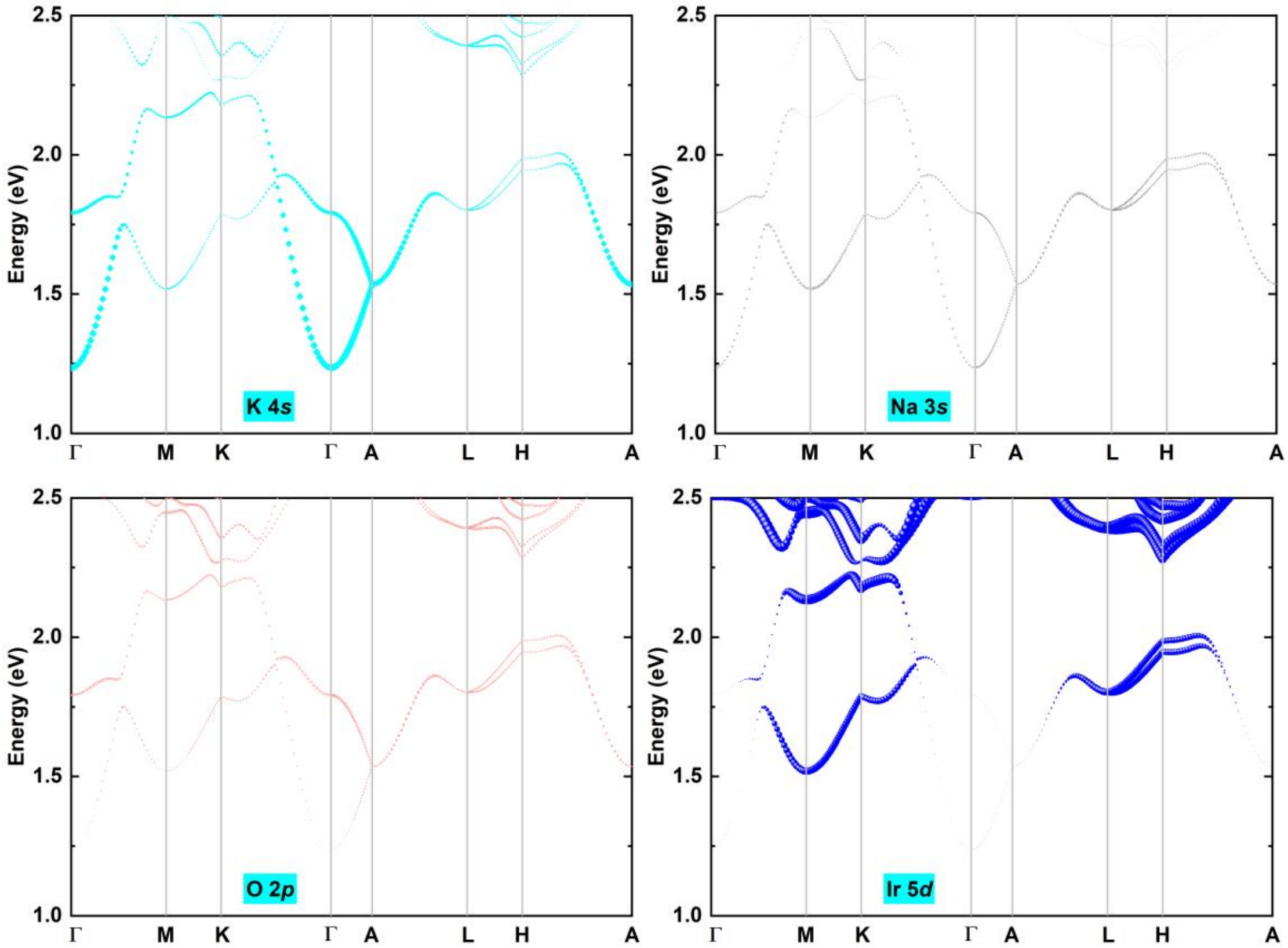

**Figure 6** Projected CBM of $K_{0.75}Na_{0.25}IrO_2$ calculated within optPBE-vdW + *U* with SOC, which are projected onto K 4*s*, Na 3*s*, O 2*p* and Ir 5*d* states, denoted by cyan diamond, black circle, red hollow balls and blue solid balls. The size of the symbols is proportional to the contribution from the corresponding elements.

As shown in **Figure 7**, we can gain further insight into the nature of the NFE states by looking at the composition of the charge densities at Γ point for the CBM. The charge densities are delocalized, where the charge densities are not localized around alkali metal ions but are distributed within the whole interstitial region between two $[IrO_2]^-$ layers. The NFE states have been proposed as ideal electron transport channels without nuclear scattering [49]. They are originally proposed to exist in graphite and alkali graphite intercalation compounds, displaying free-electron character parallel to the graphitic layers [50]. NFE states have been widely discovered in other carbon-based materials (such as graphene, nanotubes, and fullerenes [48,51-53]) as well as two-dimensional transition metal carbides and nitrides (MXenes [46], hexagonal boron nitride [41], and $Ca_2N$ monolayer [49]). The NFE states are usually unoccupied and located at high energy region, and unsuitable to be used as transport channel [46]. Fortunately, the NFE states in the title iridate $K_{0.75}Na_{0.25}IrO_2$ are located at the CBM, which renders them more accessible for electron transport or other applications than other known two-dimensional systems [46,49].

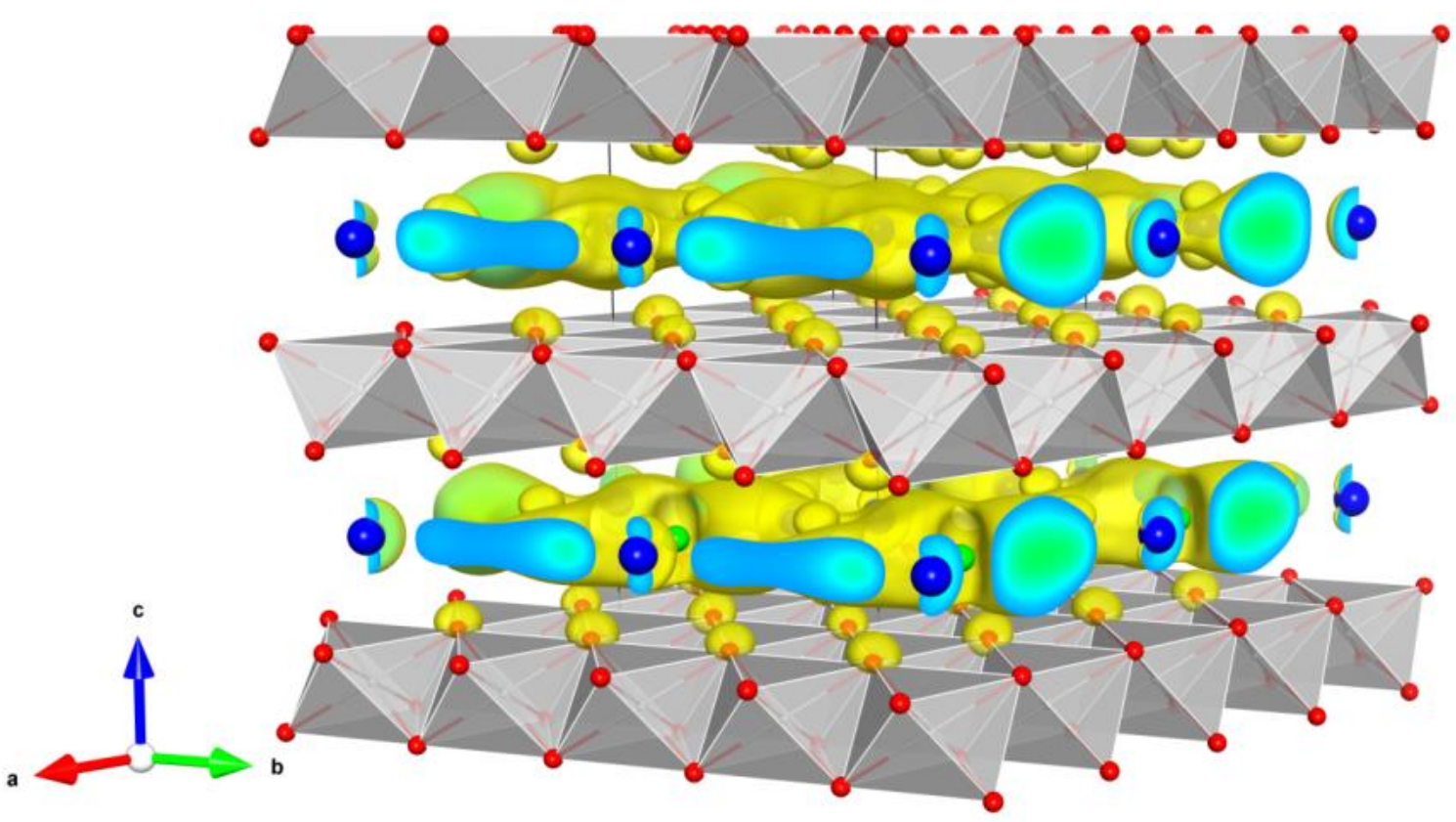

**Figure 7** Charge density maps (isosurface value is 0.008 e/ Å$^3$) of the CBM state at the Γ point for the $K_{0.75}Na_{0.25}IrO_2$ calculated within optPBE-vdW + *U*.

In addition to the fascinating in-gap states, the electronic structure of $K_{0.75}Na_{0.25}IrO_2$ also exhibits the attractive band-convergence characteristics. As shown in the zoom-in view of the band structure around Fermi level (**Figure S5** of SM [34]), the conduction bands of $K_{0.75}Na_{0.25}IrO_2$ possess two other local band-minima locating at the M point and A point beyond the global CBM of Γ point, whereas the valence band maximum (VBM) locates along the Γ-M direction accompanying with three other local extrema along the Γ-K, L-A and H-A directions, respectively. The bands are regarded as effectively converged when the band extrema of the multiple bands have no or little difference in energy, which leads to the increasing of the valley degeneracy and enhancement of electrical conductivity [54,55]. The band convergence of multi-valley conduction bands and multi-peak valence bands has been demonstrated as an effective strategy in improvement of the electrical conductivity and quality factor of thermoelectric materials [56]. In this sense, $K_{0.75}Na_{0.25}IrO_2$ could be a promising candidate of thermoelectric material with high electrical conductivity provided that the energy differences between the local band-extrema and CBM/VBM are within two times of the thermal energy at room temperature ($2k_B$T, where $k_B$ is the Boltzmann constant).

The uniaxial strain is an adequate means to manipulate the band convergence and the number of the degenerated band valleys [57]. We simulate the mechanical strain by changing the lattice constant *c* to probe the strain effect on the electronic structure of the $K_{0.75}Na_{0.25}IrO_2$. The relationship between mechanical strain and lattice constant can be expressed as: $\varepsilon = (c - c_0)/c_0$, where $c$ and $c_0$ are the lattice constant $c$ for the strained and unstrained cell, respectively. A negative (positive) value of $\varepsilon$ is defined as compressive (tensile) strain, which are realized by compressing (stretching) the crystal lattice along the $c$ axis direction. In the strain-free state, the $K_{0.75}Na_{0.25}IrO_2$ is an indirect bandgap semiconductor. The evolution of the band structure along with the applied strain is summarized in **Figure S6** of SM [34]. As shown in **Figure 8(a)**, the CBM, VBM and band gap linearly vary with the applied strain. The CBM and the VBM increase upon the increasing compressive strain ($\varepsilon < 0$), which decrease along with the tensile strain ($\varepsilon > 0$). When subjected to tensile stress along $c$-axis, except for leading to band gap shrinkage, the tensile strain plays minor impact on the dispersion of the conduction bands and the in-gap states (detailed band structures are shown in **Figure S7** of SM [34]). By contrast, as the compressive strain increases, the conduction bands near the Γ point move up and the bands near the M point move down, eventually causing the CBM to be converted from the $s$ states of alkali metal ions to Ir-5$d$ states (detailed band structures are shown in **Figure S8** of SM [34]), and resulting in band gap enlargement.

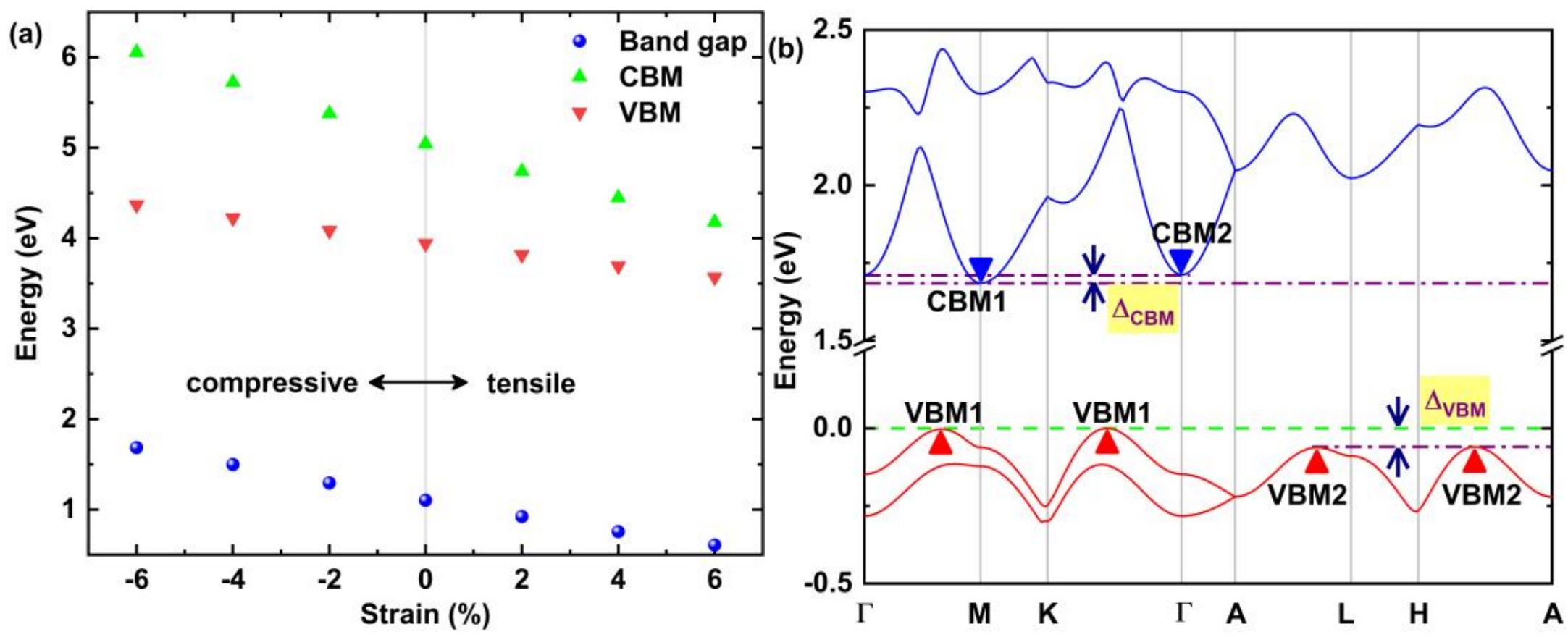

**Figure 8** (a) The evolutions of the CBM, VBM, and band gap along with the applied strain calculated with optPBE-vdW + $U$. (b) A zoom-in view of band convergence and multiple valleys achieved by 6% compressive strain.

As shown in **Figure 8(b)**, the local valence-band extremum along the Γ-K direction is elevated and becomes degenerate with the VBM along the Γ-M direction when the compressive strain increases up to 6% (which are denoted as VBM1). In addition, the local valence-band extremum along the H-A direction becomes degenerate with the extremum along the L-A direction (which are denoted as VBM2). The energy difference between VBM1 and VBM2 is about 62 meV, which is close to $2k_B$T (~52 meV). More interestingly, the locating of the global CBM has changed from the Γ point to M point along with the compression, which leads to two conduction band valleys. The energy difference between these two valleys of CBM1 and CBM2 is only 25 meV, which is far less than $2k_BT$. According to the evolution tendency of band structure under mechanical strain, the concomitant variation of the band edge may lead to completely degenerate CBM. The convergence of the band extrema has led to prominent increase of the valley degeneracy, which are highly likely to make contribution to the electrical conductivity and thermoelectric quality factor of $K_{0.75}Na_{0.25}IrO_2$ [54-57]. The above theoretical results indicate that the electronic structure of $K_{0.75}Na_{0.25}IrO_2$ could be effectively modulated by strain, and strain-induced band convergence could be an effective method to tune the thermoelectric performance of $K_{0.75}Na_{0.25}IrO_2$.

## IV. CONCLUSIONS

In summary, the first-principles DFT calculations and simulations have been performed to explore the intrinsic electronic structure of the layered iridates $K_{0.75}Na_{0.25}IrO_2$. The preferred-occupied positions of the alkali metal ions are theoretically reproduced from an energetic viewpoint. The crystal-field splitting is sufficient to rationalize the insulating nature, and the layered iridates $K_{0.75}Na_{0.25}IrO_2$ can be classified as band insulator. The SOC interactions play critical role in the band structure, resulting in NM $J_{\mathrm{eff}} = 0$ states for the trivalent $Ir^{3+}$ ($5d^6$) ions. Intriguingly, the electronic structure calculations reveal the presence of in-gap states located below the conduction bands, which are responsible for the experimentally observed abnormal low activation energy in $K_{0.75}Na_{0.25}IrO_2$. In addition, the in-gap states with parabolic energy dispersions exhibit NFE characteristics, which primarily come from the inherent alkali metal ions rather than the impurities or lattice defects. Furthermore, the band structure and band edge exhibit excellent tunability under mechanical strains. Particularly, the VBM and CBM are effectively converged by the uniaxial compression, which gives rise to significant increase of the valley degeneracy and will significantly contribute to the electrical conductivity of $K_{0.75}Na_{0.25}IrO_2$. Present theoretical results provide new insights into the unique electronic structures of the layered structure iridates, which implies $K_{0.75}Na_{0.25}IrO_2$ is a potential material for future nanoelectronic and thermoelectric applications, waiting for further experimental verifications.

## ACKNOWLEDGMENTS

We gratefully acknowledge the financial support by the National Natural Science Foundation of China (No. 11864008, 12264011), Guangxi Natural Science Foundation (No.

2018GXNSFAA138185 and AD19110081). The computational work in this research was carried out at Shanxi Supercomputing Center, and the calculations were performed on TianHe-2.